\documentclass[onecolumn,amsmath,amssymb,floatfix,10pt,prd,superscriptaddress,nofootinbib]{revtex4}

\usepackage{graphicx,epsfig}
\usepackage{amsmath,mathrsfs,amsfonts}
\usepackage {amssymb}
\usepackage {longtable}
\usepackage{multirow}
 \usepackage{enumerate}

\usepackage{bm}
\usepackage{amsfonts}
\usepackage{subfigure}
\usepackage{color}
\usepackage{relsize}
\usepackage[utf8]{inputenc}
\usepackage{hyperref}
\usepackage{orcidlink}
\usepackage{float}

\newcommand{\be}{\begin{equation}}
\newcommand{\ee}{\end{equation}}
\newcommand{\bea}{\begin{eqnarray}}
\newcommand{\eea}{\end{eqnarray}}

\begin{document}


\title{ Cosmological implications of Bumblebee theory on an FLRW background}



\author{Manuel Gonzalez-Espinoza
 \orcidlink{0000-0003-0961-8029}}
\email{manuel.gonzalez@pucv.cl}
\affiliation{
Instituto de F\'{\i}sica, Pontificia Universidad Cat\'olica de 
Valpara\'{\i}so, 
Casilla 4950, Valpara\'{\i}so, Chile.}


\author{Grigorios Panotopoulos
 \orcidlink{0000-0003-1449-9108}}
\email{grigorios.panotopoulos@ufrontera.cl}
\affiliation{
Departamento de Ciencias F{\'i}sicas, Universidad de La Frontera, Casilla 54-D, 4811186 Temuco, Chile.}

\author{Francisco Tello-Ortiz
 \orcidlink{0000-0002-7104-5746}}
\email{francisco.tello@ufrontera.cl}
\affiliation{
Departamento de Ciencias F{\'i}sicas, Universidad de La Frontera, Casilla 54-D, 4811186 Temuco, Chile.}

\begin{abstract}
We investigate some cosmological implications at the background level of the Bumblebee model. The phase-space, the critical points, and their stability are analyzed in detail, applying well-established dynamical system techniques. In addition, upon comparison to the available supernovae and Pantheon+SHOES data, the best-fit numerical value of the unique free parameter of the model is determined. We present graphical representations of key cosmological quantities as functions of redshift, including the deceleration parameter and the dark energy equation-of-state parameter. The statefinders and the age of the Universe are also computed. Finally, a comparison to the $\Lambda$CDM model is made as well. 
\end{abstract}
\maketitle

\section{Introduction}\label{Introduction}

The theory of general relativity (GR) has been a foundational paradigm in the description of the Universe's dynamics at cosmological scales. However, recent observations have cast doubt on the validity of this paradigm in certain regimes, thereby highlighting the necessity for extensions or modifications to Einstein's theory. Among the most salient issues are the nature of dark energy (DE) and dark matter (DM), as well as the potential for Lorentz symmetry violations at fundamental scales \cite{SN2,SN1}. In this context, the Bumblebee theory emerges as a natural extension of GR, introducing a vector field with a non-zero vacuum expectation value, thereby spontaneously breaking both Lorentz diffeomorphism symmetries \cite{Kostelecky:1988zi,Kostelecky:1989jw,Bluhm:2004ep,bluhm2005,Bluhm:2005uj,Bluhm:2007bd,Filho:2022yrk,Delhom:2022xfo,Delhom:2019wcm,Lambiase:2023zeo,Gullu:2020qzu,Li:2020dln,Maluf:2020kgf,Maluf:2014dpa,Casana:2017jkc,Bertolami:2005bh,Capozziello:2023tbo,Capozziello:2023rfv}. {This theory has been studied in various areas of gravitation and cosmology, providing a framework to explore modified gravity effects that could be detectable in the evolution of the Universe \cite{Bluhm:2006im}. However, this symmetry breaking is predominant at the high-energy scales, that is, at the early Universe's epoch \cite{Kostelecky:1988zi,Mattingly:2005re,Liberati2013CQG,Carroll:2004ai,Kanno:2006ty,Barrow2012PRD,Kostelecky:2011qz}. Therefore, although this theory incorporates violation of relevant symmetries for the gravitational field, it does not preclude the cosmological principle assumption at all. Recent works along these lines have demonstrated that Lorentz-violating fields can be consistently embedded in Friedmann-Lemaître-Robertson-Walker (FLRW) cosmologies, enabling studies of primordial gravitational waves \cite{Khodadi:2025wuw}, cosmic microwave background (CBM) polarizations \cite{Kostelecky:2001mb,Kostelecky:2007zz,Caloni:2022kwp,Fairbairn:2014kda,Xia:2009ah,Yunes:2016jcc,Mirshekari:2011yq,Audren:2013dwa} and constrains on Lorentz violation  with big bang nucleosynthesis and gravitational baryogenesis \cite{Khodadi:2022mzt} without spoiling large-scale isotropy\footnote{For other applications of Bumblebee gravity theory, for example black hole solutions and related subjects see for instance \cite{AraujoFilho:2024iox,AraujoFilho:2024ykw,AraujoFilho:2025hkm,Shi:2025plr,Heidari:2024bvd,Filho:2023etf,Khodadi:2021owg,Khodadi:2022dff}}. This perspective highlights the possibility of a compatibility between Lorentz violation and the cosmological principle. Nevertheless, Bumblebee contributions could still damage the cosmological principle through non-minimal coupling to the gravity sector. In this regard, some recent studies have been carried out in the context of anisotropic cosmological models such as those based on Bianchi-I \cite{Sarmah:2024xwx} or Kasner metrics \cite{Neves:2022qyb}. Despite this fact, it is also possible to construct isotropic scenarios consistent with an FLRW background. In particular, statistical isotropy can be preserved through cosmic vector triads or by restricting the Bumblebee vacuum expectation value to be time-like \cite{Armendariz-Picon:2004say,Kostelecky:2003fs}.  }
In this way, the Bumblebee field may induce contributions to the effective equation of state, offering scenarios in which cosmic acceleration arises without the need for an explicit cosmological constant, although such behavior strongly depends on the potential and coupling choices \cite{Capelo:2015ipa,Esposito-Farese:2009wbc}.

{
 It is important to highlight that other theories that incorporate Lorentz breaking in gravity include the Einstein-Æther model \cite{Jacobson:2000xp} and Ho\v{r}ava-Lifshitz gravity \cite{Horava:2009uw,Blas:2009qj}. In this case, the Einstein-Æther theory introduces a dynamical vector field that selects a preferred reference frame at each point in spacetime, which can modify the propagation of gravitational waves and affect cosmological evolution. In contrast, Ho\v{r}ava-Lifshitz gravity aims to formulate a complete ultraviolet expression of quantum gravity by explicitly breaking Lorentz symmetry at high energies, thereby allowing the presence of high-derivative terms that enhance the renormalizability of the theory\footnote{It should be pointed out that at low energy level these theories are compatible \cite{Jacobson:2013xta}.}. However, compared to these theories, the Bumblebee model offers a conceptual advantage by preserving a covariant framework and allowing Lorentz violation to occur spontaneously rather than explicitly. In particular, the modified Friedmann equations in these models have the potential to generate observable effects in data from type Ia supernovae (SNe Ia), baryon acoustic oscillations (BAO), and the CMB. } {Although the Bumblebee theory has a richer field of research, it should be noted that Einstein-Æther, Ho\v{r}ava-Lifshitz and Bumblebee theories have been strongly constrained by observational data from neutron stars in \cite{Barausse:2019yuk,EmirGumrukcuoglu:2017cfa,Gong:2018vbo} and \cite{Ji:2024aeg}, respectively.} 

 On the other hand, there are other modified gravity theories that include vector fields as a DE source. For example, in \cite{BeltranJimenez:2008iye,BeltranJimenez:2009xus} was proposed a cosmic vector for dark energy, incorporating a dynamical vector field $A_{\mu}$ non-minimally coupled to gravity and also the so-called generalized multi-Proca fields \cite{Allys:2016kbq,BeltranJimenez:2016afo,Garcia-Serna:2025dhk,Martinez:2024gsj,GallegoCadavid:2020dho,Gomez:2019tbj,Gomez:2020sfz}. When comparing Bumblebee theory with the former, the main difference lies in the fact that Bumblebee fields cannot be time-like because they become kinematic degrees of freedom. Moreover, coupling constants in the Bumblebee theory are not fixed from the very beginning; therefore, this allows us to further constrain the Bumblebee theory using phenomenological data. Now, in comparison with Proca fields, the main contrast is that Proca fields do not induce Lorentz symmetry breaking. So, multi-Proca excels as a minimal, flexible framework for dark-energy phenomenology. Then we have different frameworks to test how dark energy can be represented by a vector.

In the context of modified gravity, the Friedmann equations take a nonlinear and highly coupled form. As a consequence, the numerical evolution of the system becomes extremely sensitive to initial conditions and parameter choices. This sensitivity poses a significant challenge for extracting meaningful cosmological behavior, especially in the absence of analytical solutions. Therefore, dynamical systems analysis provides a powerful complementary tool in this setting. Rather than focusing on specific initial conditions, this method allows us to explore the asymptotic behavior of the system across the entire parameter space. Specifically, we can determine under which conditions the cosmological solutions naturally evolve toward accelerating, expanding, or static universes \cite{Copeland:2006wr, Leon:2009rc, Rodriguez-Benites:2023otm, Gonzalez-Espinoza:2022hui}. In the case of the Bumblebee model, we find that only certain subsets of the parameter space yield late-time accelerated expansion compatible with observations. This insight helps reduce the effective volume of the parameter space, making subsequent statistical analyses (e.g., $\chi^2$ fits to observational data) more efficient and physically motivated.

On the other hand, current cosmological data indicate that the present abundances of dark energy and non-relativistic matter are of the same order of magnitude, despite the fact that the energy densities of the two fluid components evolve with time very differently. This is the coincidence or the ``why now" problem of dark energy. In cosmological models where the present Universe is realized as a finite point during the cosmic evolution, the answer to the coincidence question ``why it is that today $\Omega_{m0}$ and $\Omega_{DE,0}$ are of the same order of magnitude'' relies on an appropriate choice of special initial conditions. On the contrary, in a scenario in which the present Universe is in its asymptotic era (namely, close to a fixed point), the answer to the above question reduces to an appropriate choice of the numerical values of the free parameters of the model. Clearly, since we do not like unnatural initial conditions, the second scenario is more favorable than the first one. Therefore, given a certain cosmological model, it is interesting to see whether or not there is a stable critical point (i.e., an attractor) that corresponds to acceleration with an $\Omega_{m0}$ in the range [0,1].

The present study aims to analyze the cosmological evolution in the context of Bumblebee theory on an FLRW background by conducting a detailed study of the dynamical systems associated with its field equations. The stability of critical points will be investigated, exploring the admissible range for the coupling constant of the theory. The cosmic triad formulation will be implemented to justify the use of an FLRW background and its deviations from previous models. Finally, comparisons with observational data, in particular with data from type Ia supernovae, will be made to assess the model's ability to reproduce current cosmological trends. This analysis will facilitate a multifaceted evaluation of the model's performance, including its capacity to reproduce the observed cosmological trends. Furthermore, in the present case, the so-called statefinder parameters $r, s$  \cite{Sahni:2002fz,Alam:2003sc} are also studied. 
Those parameters may be computed within a given model; their values can be extracted from future observations \cite{SNAP1,SNAP2} and can be very different from one model to another, even if the two expansion histories are very similar. The statefinder diagnostic has been applied to several DE models; see, e.g. \cite{diagnostics1,diagnostics2,diagnostics3,diagnostics4,diagnostics5} and references therein. {The novelty of the present work lies in combining a ``cosmic triad'' configuration with a non-minimal Ricci–Bumblebee interaction and studying its cosmological consequences using dynamical system techniques. This allows us to identify the viable parameter space of the model and confront its predictions with the latest Pantheon+ supernova observations.} 
{In addition to the formal dynamical analysis, the present work aims at clarifying the physical role played by the Ricci–Bumblebee interaction in cosmology. In particular, we show that even the minimal sector of the theory characterized by a single coupling parameter $\alpha$ leads to a non-trivial cosmological dynamics. The dynamical system analysis reveals that the model naturally reproduces the standard sequence of cosmological epochs, namely a radiation-dominated phase followed by a matter-like saddle point and finally a late-time accelerated attractor.} 
{
Furthermore, the stability analysis allows us to constrain the viable interval of the coupling parameter, $1/6<\alpha<1/2$, for which the late-time attractor corresponds to an accelerated expansion phase. This provides a clear physical interpretation of the Ricci–Bumblebee interaction as a possible mechanism driving late-time cosmic acceleration. Finally, by confronting the theoretical predictions with the Pantheon+SH0ES supernova data, we obtain observational bounds on the parameter $\alpha$, showing that the model can reproduce the observed expansion history while remaining distinguishable from the $\Lambda$CDM scenario through the statefinder diagnostics.}

The present article is organized as follows: Section \ref{bum_action} is devoted to the presentation of the Bumblebee model, Section \ref{cosmo_dyna} presents the general setting to address the problem, that is, the adapted Friedmann equations, the cosmological parameters and also the results provided by the dynamical system approach (phase space analysis and stability of critical points) analysis, comparison with observational SN data and statefinder diagnostic. Finally, Section \ref{Concluding_Remarks} closes the work, highlighting the main results, challenges, and outlook.

\section{Bumblebee Action}\label{bum_action}

{The most general and simplest action considering a single Bumblebee field $B_{\mu}$ coupled to gravity is given by \cite{Bluhm:2007bd}
\begin{equation}
S_B  =\int  d^4 x \sqrt{-g}\mathcal{L}_B 
 =\int d^4 x \sqrt{-g} \left(\mathcal{L}_g+\mathcal{L}_{g B}+\mathcal{L}_{\mathrm{K}}+\mathcal{L}_V+\mathcal{L}_{\mathrm{M}}\right),
\end{equation}
where (from left to right) one has the pure gravitational Lagrangian density (commonly the Einstein-Hilbert term with/without cosmological constant), the gravity-Bumblebee coupling, the Lorentz violating Lagrangian density, and the matter Lagrangian density. Specifically, the Bumblebee Lagrangian density considers only quadratic terms in $B_{\mu}$ and involves at most no more than two derivatives is  
\begin{equation}\label{bum}
\begin{aligned}
\mathcal{L}_B= & \frac{1}{16 \pi G} (R-2 \Lambda)+\sigma_1  B^\mu B^\nu R_{\mu \nu}+\sigma_2  B^\mu B_\mu R  -\frac{1}{4} \tau_1  B^{\mu \nu} B_{\mu \nu}+\frac{1}{2} \tau_2  D_\mu B_\nu D^\mu B^\nu  +\frac{1}{2} \tau_3  D_\mu B^\mu D_\nu B^\nu- V+\mathcal{L}_{\mathrm{M}},
\end{aligned}
\end{equation}
with $G$ being Newton's gravitational constant, and the field strength tensor $B_{\mu\nu}$ is defined as 
\begin{equation}
B_{\mu \nu}=\partial_\mu B_\nu-\partial_\nu B_\mu.
\end{equation}
As can be observed, there are five real coupling constants $\sigma_1, \sigma_2, \tau_1, \tau_2$ and $\tau_3$ in the Lagrangian density (\ref{bum}) that are not independent of each other.\footnote{To be more precise, the action (\ref{bum}) can contain a boundary term (not affecting the field equations)
\begin{equation}
\int d^4 x\sqrt{-g}\left( B^\mu B^\nu R_{\mu \nu}-\frac{1}{2}  B^{\mu \nu} B_{\mu \nu}  + D_\mu B_\nu D^\mu B^\nu- D_\mu B^\mu D_\nu B^\nu\right)=0
\end{equation}
linking parameters $\sigma_1, \tau_1, \tau_2, \tau_3$. } 
On the other hand, the potential $V$ responsible for the spontaneous Lorentz symmetry breaking and  diffeomorphism rupture when coupling with gravity, has no derivative couplings and is
formed from scalar combinations of the Bumblebee field $B_{\mu}$ and the metric tensor $g_{\mu\nu}$, that is
\begin{equation}
    V=V(X), \quad X=B_\mu g^{\mu \nu} B_\nu \pm b^2,
\end{equation}
where $b$ is a positive real constant. The vacuum is determined by the single
condition
\begin{equation}
X=B_\mu g^{\mu \nu} B_\nu \pm b^2=0,
\end{equation}
producing $V(X)=V'(X)=0$ and the fields $B_{\mu}$ and $g_{\mu\nu}$ acquire vacuum values 
\begin{equation}
B_\mu \rightarrow\left\langle B_\mu\right\rangle=b_\mu, \quad g_{\mu \nu} \rightarrow\left\langle g_{\mu \nu}\right\rangle.
\end{equation}
The non zero value of $b_{\mu}$, which obeys $b_\mu\left\langle g^{\mu \nu}\right\rangle b_\nu=\mp b^2$, spontaneously breaks both Lorentz and diffeomorphism
symmetry. This, in turn, affects the dynamics of other fields that interact with the Bumblebee field. Despite these modifications, the fundamental geometric properties and conservation laws remain consistent with the pseudo-Riemannian structure commonly used in general relativity \cite{kostelecky2004}. It is worth mentioning that the choice of potential in Eq. (\ref{bum}) can also have an incidence for the parameters $\sigma_1, \sigma_2, \tau_1, \tau_2$ and $\tau_3$ \cite{Bluhm:2007bd}. Furthermore, the form of the potential $V(X)$ is very important to ensure that the vacuum expectation value $\langle b \rangle$ remains constant, i.e., unchanged. This means that the field $b$ is fixed at a specific value and does not experience fluctuations around it. In this type of potential, the system is restricted to a single state or configuration, which limits its dynamics. Some possible forms adopted for the potential $V(X)$ are listed in table \ref{tab:Bumblebee_potentials}, where the first one, the so-called linear Lagrange multiplier, will be used in this case, to ensure the above appointments. 
\begin{table}[H]
    \centering
    \begin{tabular}{|c|c|c|c|c|}
        \hline
        \textbf{Potential} & \( V(B_\mu B^\mu) \) & \( V(b) = 0 \)? & \textbf{Strict Constraint?} & \textbf{Allows Fluctuations?} \\
        \hline
        \textbf{Lagrange Multiplier} & \( \lambda (B_\mu B^\mu \pm b^2) \) &  Yes &  Yes &  No \\
        \hline
        \textbf{Quadratic Soft} & \( \lambda (B_\mu B^\mu \pm b^2)^2 \) &  Yes &  No &  Yes \\
        \hline
        \textbf{Higher Order} & \( \lambda (B_\mu B^\mu \pm b^2)^n, \, n>1 \) &  Yes &  No &  Yes \\
        \hline
        \textbf{Nonlinear Corrections} & \( \lambda (B_\mu B^\mu \pm b^2)^2 + \alpha (B_\mu B^\mu \pm b^2)^3 \) &  Yes & No &  Yes \\
        \hline
    \end{tabular}
    \caption{Comparison of different potential choices for Bumblebee models.}
    \label{tab:Bumblebee_potentials}
\end{table}}
{ Since the term $B^{\mu}B^{\nu}R_{\mu\nu}$ induces anisotropies in the effective energy-momentum tensor, transgressing the cosmological principle since the isotropy is broken, we adopt the proposal of the cosmic triad \cite{Bento:1992wy,Armendariz-Picon:2004say,Wei:2006tn}, which introduces three mutually orthogonal vectors in space. By averaging in the three directions, the effective isotropy can be recovered on large scales, allowing the use of an FLRW background. Then, the Bumblebee field is expressed as 
\begin{equation}
    B_{\mu}\rightarrow B^{a}_{\mu},
\end{equation}
being $a$ the cosmic triad index. This method introduces three mutually orthogonal spatial directions. So, at large scales, one can average on these directions and, in this way, recover the isotropy of spacetime.}  Under this assumption, the field strength and the conditions over the potential $V(X)$ are written as   
\begin{equation}
{B^a}_{\mu\nu}=\partial_\mu B^a_\nu-\partial_\nu B^a_\mu,
\end{equation}
and 
\begin{equation}
X = b^{a\mu} \langle g_{\mu\nu} \rangle b^{a\nu} \pm b^{a\mu} b^a{}_{\mu} = 0
\end{equation}
implying
\begin{equation}
    \left. V\left(B^{a\mu} B^a{}_{\mu} \pm b^{a\mu} b^a{}_{\mu} \right) \right|_{B^{a\mu} = b^{a\mu}} = 0, \quad \text{and} \quad \left. V^{\prime} \left(B^{a\mu} B^a{}_{\mu} \pm b^{a\mu} b^a{}_{\mu} \right) \right|_{B^{a\mu} = b^{a\mu}} = 0,
\end{equation}
where the vacuum expectation values of the fields are given by  
\begin{equation}
B^{a\mu} \to \langle B^{a\mu} \rangle = b^{a\mu}, \quad g_{\mu\nu} \to \langle g_{\mu\nu} \rangle. \label{vacuum_cond}
\end{equation}

It is important to stress that the triad configuration introduces genuine new physics compared to the case where only the temporal component of the vector field is considered. In the latter case, the field strength vanishes, 
$B_{\mu\nu}=0$, so that the kinetic term $\tfrac{1}{4}{B^a}_{\mu\nu}{B^a}^{\mu\nu}$ is identically zero and the system lacks any intrinsic dynamics, effectively reducing to a trivial modification of the background. By contrast, with the cosmic triad, the kinetic term remains nonvanishing, allowing the vector sector to evolve dynamically and contribute nontrivially to the cosmological equations.

Taking into account these facts, the action of the Bumblebee theory can be cast as
\bea
&S=&\int \text{d}^{4}{x} \sqrt{-g}\Bigg[\frac{R}{2\kappa^2} -\sum_{a=1}^3 \frac{1}{4}{B^a}_{\mu\nu}{B^a}^{\mu\nu}+ {\alpha}\sum_{a=1}^3 B^{a \ \mu} B^{a \ \nu} R_{\mu \nu}-\sum_{a=1}^3 V\left(B^{a \ \mu} B^{a}_{\ \ \mu} \pm b^2\right)\Bigg]+ S_{m}, \label{action00}
\eea
where $\kappa^2 = 8\pi G$ is the gravitational coupling.
Notice that we have considered a particular case of action (\ref{bum}), that is, the case $\tau_{2}=\tau_{3}=0$, $\sigma_{2}=0$, $\tau_{1}=1$ and $\sigma_{1}=\alpha$. 
Also, we set $\Lambda=0$, since it is expected that the Bumblebee field acts as an effective cosmological constant, producing the Universe acceleration. It is remarkable to note that this action corresponds to the family of the Bumblebee model given in \cite{Maluf:2021lwh,Jesus:2019nwi}. {It is worth mentioning that all the restrictions considered here are also useful in view of recent results showing that the most general marginal Bumblebee vector-tensor cosmology faces a no-go theorem at the level of linear cosmological perturbations \cite{vandeBruck:2025aaa}.}

Next, the field equations that govern the gravitational sector in Bumblebee gravity are obtained by taking variations of the action \eqref{action00} with respect to the metric \( g_{\mu\nu} \). This process results in the modified Einstein field equations

\begin{align}
G_{\mu\nu} &= \kappa^2 \left( T^B_{\mu\nu} + T_{\mu\nu} \right) \notag \\
&= \kappa^2 \sum_{a=1}^3 \left[ 2V' B^{a}{}_{\mu} B^{a}{}_{\nu} + B^{a\alpha}{}_{\mu} B^{a}{}_{\nu\alpha} - \left( V + \frac{1}{4} B^{a\alpha\beta} B^{a}{}_{\alpha\beta} \right) g_{\mu\nu} \right] \notag \\
&\quad + 2 \kappa^2 \alpha  \sum_{a=1}^3 \bigg[ \frac{1}{2} B^{a\alpha} B^{a\beta} R_{\alpha\beta} g_{\mu\nu} - B^{a}{}_{\mu} B^{a\alpha} R_{\alpha\nu} - B^{a}{}_{\nu} B^{a\alpha} R_{\alpha\mu} \notag \\
&\quad + \frac{1}{2} \nabla^\alpha \nabla_\mu (B^{a}{}_{\alpha} B^{a}{}_{\nu}) + \frac{1}{2} \nabla^\alpha \nabla_\nu (B^{a}{}_{\alpha} B^{a}{}_{\mu}) \notag \\
&\quad - \frac{1}{2} \nabla^2 (B^{a}{}_{\mu} B^{a}{}_{\nu}) - \frac{1}{2} g_{\mu\nu} \nabla^\alpha \nabla^\beta (B^{a}{}_{\alpha} B^{a}{}_{\beta}) \bigg] + \kappa^2 T_{\mu\nu}. \label{background0}
\end{align}

The Einstein tensor is denoted as \( G_{\mu\nu} \), and the prime notation indicates differentiation with respect to the argument of the potential. The symbols \( T^B_{\mu\nu} \) and \( T_{\mu\nu} \) represent the energy-momentum tensors of the Bumblebee field and the matter-radiation content, respectively. 

To address the highly non-linear set of equations \eqref{background0}, we need to impose a specific metric ansatz such as the cosmic triad considered in this study—along with an appropriate choice for the Bumblebee field (see next section \ref{cosmo_dyna}). The exact form of the potential \( V \) is not critical to our approach, as we focus our analysis on the vacuum regime where \( V = V' = 0 \). 

Then, the action \eqref{action00} leads to a field equation for the Bumblebee field itself. By varying the action with respect to \( B^{a\mu} \), we obtain the corresponding field equation

\begin{equation}
\nabla^\mu B^{a}{}_{\mu\nu} = 2 \left( V' B^{a}{}_{\nu} - \alpha B^{a\mu} R_{\mu\nu} \right),
\end{equation}

which governs the dynamics of the Bumblebee field $B^{a\mu}$. For simplicity, we assume there is no direct coupling between the matter sector and the Bumblebee field.

{Finally, it is worth noting that the reduced action \eqref{action00} arises directly from the considerations discussed above. The most general quadratic Bumblebee action contains several coupling terms, some of which are related through boundary identities and therefore are not independent. In addition, we adopt a Lagrange multiplier potential and restrict the analysis to the vacuum regime, where $V = V' = 0$. In this situation, the potential does not contribute to the background dynamics, and related cosmology is mainly determined by the vector field's kinetic sector and its non-minimal coupling to curvature. For this reason, we focus on the sector described by action \eqref{action00}, which is controlled by a single coupling parameter $\alpha$. This term captures the Ricci-Bumblebee interaction while keeping the system analytically manageable. As shown below, even this simplified model shows a nontrivial phase-space structure, including radiation- and matter-dominated saddle points, as well as a late-time attractor compatible with late-time observation data. The simplified sector considered here should therefore be viewed as a controlled setup that allows us to isolate the cosmological role of the Ricci-Bumblebee interaction.}

\section{Cosmological dynamics}\label{cosmo_dyna}

{In this section, we present the field equations  (\ref{background0}) adapted to an FLRW spacetime, leading to the modified Friedmann equations. Also, the effective equation of state (EoS) parameter is defined, and cosmological quantifiers such as density parameters and statefinder, deceleration $q$, jerk $r$, and snap $s$ parameters are introduced. Next, the phase space analysis through a dynamical system approach is implemented in order to study the stability of critical points. Finally, using the $\chi^{2}$-square method along with SN data, the best value for the constant coupling gravity-Bumblebee is determined.}


\subsection{Friedmann equations and cosmological parameters}

{To explore cosmological consequences within the gravity-Bumblebee scenario, we adopt the well-known FLRW spacetime, being described by the following line element}

\begin{equation}
ds^2=-dt^2+a^2\,\delta_{ij} dx^i dx^j \,,
\label{FRWMetric}
\end{equation}
where $a=a(t)$ is the scale factor, a function of cosmic time $t$ and {we consider a spatially flat case $k=0$}. In addition, a Bumblebee field as $B^a_\mu=\delta^a_\mu B(t)a(t)$ \cite{Armendariz-Picon:2004say}, {to preserve the cosmological principle.}
Therefore, the modified Friedmann equations are 
\bea
\label{MFreq1}
  \frac{3 H^2}{\kappa ^2} &=& \dfrac{3}{2} \left(\dot{B}+ H B \right)^2 - 6 \alpha B \dot{B} H+\rho _m+\rho _r,\label{00}\\
 \frac{-2 \dot{H}}{\kappa ^2} &=& 2 \left(\dot{B}+ H B \right)^2 - 6 \alpha B^2 H^2 - 6 \alpha B \dot{B} H + 2\alpha \dot{B}^2 - 2\alpha B^2 \dot{H} + 2\alpha B \ddot{B}
+\rho _m+\frac{4 \rho _r}{3},\label{ii}
\eea and the motion equation for $B^a_\mu$ given by 
\bea
&& \dfrac{d (\dot{B} + B H)}{d t} + 2 H (\dot{B} + B H) - 2 \alpha B \dot{H} - 6 \alpha B H^2  =0.
\label{MFreq2}
\eea 

Now, introducing the following effective quantities 
\bea
\label{rhode}
\rho_{de}&= & \dfrac{3}{2} \left(\dot{B}+ H B \right)^2 - 6 \alpha B \dot{B} H,\\
 p_{de}&= & \dfrac{1}{2} \left(\dot{B}+ H B \right)^2 - 6 \alpha B^2 H^2  + 2\alpha \dot{B}^2 - 2\alpha B^2 \dot{H} + 2\alpha B \ddot{B}, \nonumber\\
 \label{pde}
&&\eea 
the field equations (\ref{MFreq1}) and (\ref{00}) adopt the following form
\bea
\label{SH00}
&& \frac{3}{\kappa^2} H^2=\rho_{de}+\rho_{m}+\rho_{r},\\
&& -\frac{2}{\kappa^2} \dot{H}=\rho_{de}+p_{de}+\rho_{m}+\frac{4}{3}\rho_{r}.
\label{SHii}
\eea

Then, the effective dark energy EoS parameter is given by
\begin{equation}
w_{de}=\frac{p_{de}}{\rho _{de}},
\label{wDE1}
\end{equation}
and the effective thermodynamics quantities $\rho_{de}$ and $p_{de}$ are satisfying 
\begin{eqnarray}
\dot{\rho}_{de}+3H(\rho_{de}+p_{de})=0.
\end{eqnarray} 
This equation is consistent with the energy conservation law and the fluid evolution equations
\bea
\label{rho_m}
&& \dot{\rho}_{m}+3 H\rho_{m}=0,\quad 
\dot{\rho}_{r}+4 H\rho_{r}=0,
\label{rho_r}
\eea
{therefore, unlike other models, such as for instance \cite{Jawad4,Jawad5}, the Bumblebee model considered here is a non-interacting DE model.}

At this stage, it is useful to introduce the total EoS parameter as 
\be
w_{tot}=\frac{p_{de}+p_r}{\rho _{de}+\rho _m+\rho _r},
\label{wtot}
\ee which is related to the deceleration parameter $q$ through
\be
q=\frac{1}
{2}\left(1+3w_{tot}\right).
\label{deccelparam}
\ee Then, the acceleration of the Universe occurs for $q<0$, or equivalently for  $w_{tot}<-1/3$.

Furthermore, another useful set of cosmological parameters is the so-called standard density parameters  
\bea
&& \Omega_{m}\equiv\frac{\kappa^2 \rho_{m}}{3 H^2},\:\:\:\: \Omega_{de}\equiv\frac{\kappa^2 \rho_{de}}{3 H^2},\:\:\:\: \Omega_{r}\equiv \frac{\kappa^2 \rho_{r}}{3 H^2},
\eea which satisfies the constraint equation  
\be\label{omegadensity}
\Omega_{de}+\Omega_{m}+\Omega_{r}=1.
\ee 

{Besides the deceleration parameter $q$, also we introduce another statefinder parameters such the jerk $r$ and snap $s$, defined as follows
\be
r \equiv \frac{\dddot{a}}{a H^3}, \; \; \; \; \; s \equiv \frac{r-1}{3 (q-1/2)},
\ee }
 where for the $\Lambda$CDM model, these parameters take the values $r=1,s=0$.

For later purposes, it is useful to express these quantifiers in terms of the redshift, $z=-1+a_0/a$, with $a_0$ being the present value of the scale factor. So, the deceleration parameter and the jerk statefinder are given by the following expressions
\begin{eqnarray}
q(z) & = & -1+(1+z) \: \frac{H'(z)}{H(z)} \\
r(z) & = & (1+z) \: q'(z) + q(z) [1+2 q(z)].
\end{eqnarray}

\subsection{Phase space analysis}

Now that all the necessary cosmological parameters and equations have been introduced, we may implement the dynamical system analysis \cite{Jawad1,Jawad2,Jawad3}. To that end, it is more convenient to define the following useful dimensionless variables

\begin{eqnarray}
x =& \dfrac{\kappa  \dot{B}}{\sqrt{6} H}, \ \ \ \ \ \ y =& \dfrac{\kappa  B}{\sqrt{6} }, \ \ \ \ \ \ \ \varrho= \frac{\kappa\sqrt{\rho_r}}{\sqrt{3}H}.
\label{var}
\end{eqnarray}
From equations \eqref{MFreq1} and (\ref{omegadensity}) it is obtained
\begin{equation}
    \Omega_m  +3 x^2+6 x y+3 y^2-12 \alpha  x y+\varrho ^2= 1,
\end{equation}
and from equations \eqref{ii} and  \eqref{MFreq2}, one gets the following dynamical system 
\begin{eqnarray}
\dfrac{d x}{d N} &=& \frac{f_1(x,y,\varrho)}{24 (\alpha -1) \alpha  y^2+2},
\label{dinsyseq1}\\
\dfrac{d y}{d N} &=& x, \label{dinsyseq2}\\
\dfrac{d \varrho}{d N} &=& \frac{\varrho f_2(x,y,\varrho)}{24 (\alpha -1) \alpha  y^2+2}. \label{dinsyseq3}
\end{eqnarray}
where
{\small
\begin{eqnarray}
f_1 &=& 3 (4 \alpha +1) x^3-3 \left(8 \alpha ^2+10 \alpha -3\right) x^2 y+x \left(9 \left(8 \alpha ^2-4 \alpha +1\right) y^2+\varrho ^2-3\right) \nonumber\\
&+&y \left(-2 \alpha  \varrho ^2+6 \alpha +(3-18 \alpha ) y^2+\varrho ^2-1\right), \nonumber\\
f_2 &=& 3 (4 \alpha +1) x^2+6 (1-6 \alpha ) x y+3 \left(8 \alpha ^2-4 \alpha +1\right) y^2+\varrho ^2-1 ,
\end{eqnarray}}

Now, utilizing the above phase space variables, we can also express the density parameters as
\bea
\Omega_{de} &=& 3 x^2-12 \alpha  x y+6 x y+3 y^2,\\
\Omega_{m} &=& 1 -3 x^2+12 \alpha  x y-6 x y-3 y^2-\varrho ^2 
\eea
and the EoS parameter for the dark energy can be reformulated as 
\bea
w_{de} &=& \frac{(4 \alpha +1) x^2+2 (1-6 \alpha ) x y+y^2 \left(-4 \alpha ^2 \left(\varrho ^2-3\right)+4 \alpha  \left(\varrho ^2-2\right)+1\right)}{\left(12 (\alpha -1) \alpha  y^2+1\right) \left(3 x^2-12 \alpha  x y+6 x y+3 y^2\right)} ,\nonumber\\
&&
\eea
whereas the total EoS parameter takes the form
\bea
w_{tot}&=& \frac{3 (4 \alpha +1) x^2+6 (1-6 \alpha ) x y+3 \left(12 \alpha ^2-8 \alpha +1\right) y^2+\varrho ^2}{36 (\alpha -1) \alpha  y^2+3} \nonumber\\
\eea


\textbf{Critical points}\\

Next, we find the critical points. To do so, one imposes the conditions \( \frac{dx}{dN} = \frac{dy}{dN} = \frac{d\varrho}{dN} = 0 \) on the system ~\eqref{dinsyseq1}-\eqref{dinsyseq3}. This provides the critical points summarized in Table~\ref{table1}, while Table~\ref{table2} provides the associated cosmological parameters.

The critical point \( a_R \) corresponds to the radiation-dominated phase, characterized by \( \Omega_r = 1 \) and an EoS parameter \( \omega_{tot} = \frac{1}{3} \). The positive value of the deceleration parameter confirms that this point denotes a decelerating phase in cosmic evolution. {It is important to note that at the critical point \( a_R \), the conditions are such that \( \Omega_{r} = 1 \) and \( \Omega_{\rm de} = 0 \). This means that if our universe begins at \( a_R \), it automatically satisfies the nucleosynthesis constraint \( \Omega_{\rm de} < 0.045 \) \cite{Bean:2001wt}. This condition ensures that the dark energy sector remains negligible during the radiation-dominated epoch, which is also numerically corroborated in the next section. On the other hand, a detailed discussion of 
the possible impact on the effective number of relativistic species, while 
certainly interesting, lies beyond the scope of the present work.
}

The critical point \( b_M \) represents the matter-dominated era, where \( \Omega_m = 1 \) and \( \omega_{tot} = 0 \). The positive value of the 
deceleration parameter indicates that this point corresponds to a decelerating phase of cosmic evolution. This stage plays the expected intermediate role 
between the radiation epoch ($a_R$) and the dark energy dominated attractor  ($c$), showing that the model naturally reproduces the standard sequence of cosmic history.

For the critical point \( c \), we have \( \Omega_{de} = 1 \) and an equation of state given by
\( \omega_{de} = \omega_{tot} = \frac{1 - 6\alpha}{3 - 6\alpha}\). Depending on the value of \( \alpha \), this point can describe a phase of accelerated expansion.
{In fact, as shown in the stability analysis, the critical point $c$ corresponds to an accelerated attractor provided that the coupling constant lies in the interval $1/6 < \alpha < 1/2$. Within this range, the Universe naturally evolves toward a late-time phase of accelerated expansion dominated 
by dark energy.
}

\begin{figure}[!h]
    \centering
        \includegraphics[scale=0.2]{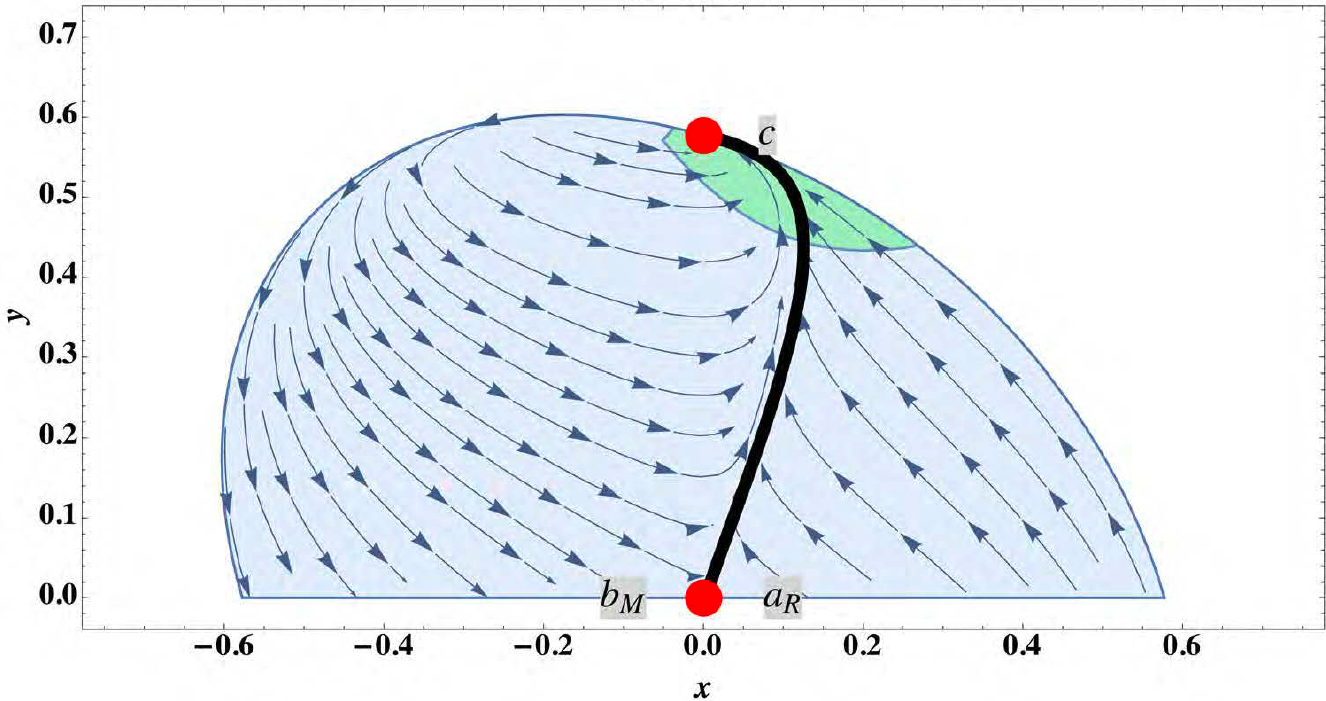}
        \caption{\scriptsize The phase space of \( x \) versus \( y \) for \(\alpha = 0.347\) is shown, where the critical points are marked as red dots with their respective labels. Each trajectory represents the evolution of the Universe for different initial conditions. In particular, the black line corresponds to a Universe initialized with \( x_i = 8.0 \times 10^{-9} \), \( y_i = 1.645 \times 10^{-3} \), and \( \varrho_i = 0.999825 \). The green region highlights the domain of accelerated expansion. {Since the dynamical system is symmetric under the transformation $y \to -y$, we restrict  the analysis to the upper half-plane ($y>0$) without loss of generality. In this way, only the three relevant critical points of Table~\ref{table1} are displayed, which are sufficient to reproduce the thermal history of the Universe.}}  
    \label{2Dphase1}
\end{figure}

\begin{figure}[!h]
    \centering
        \includegraphics[scale=0.35]{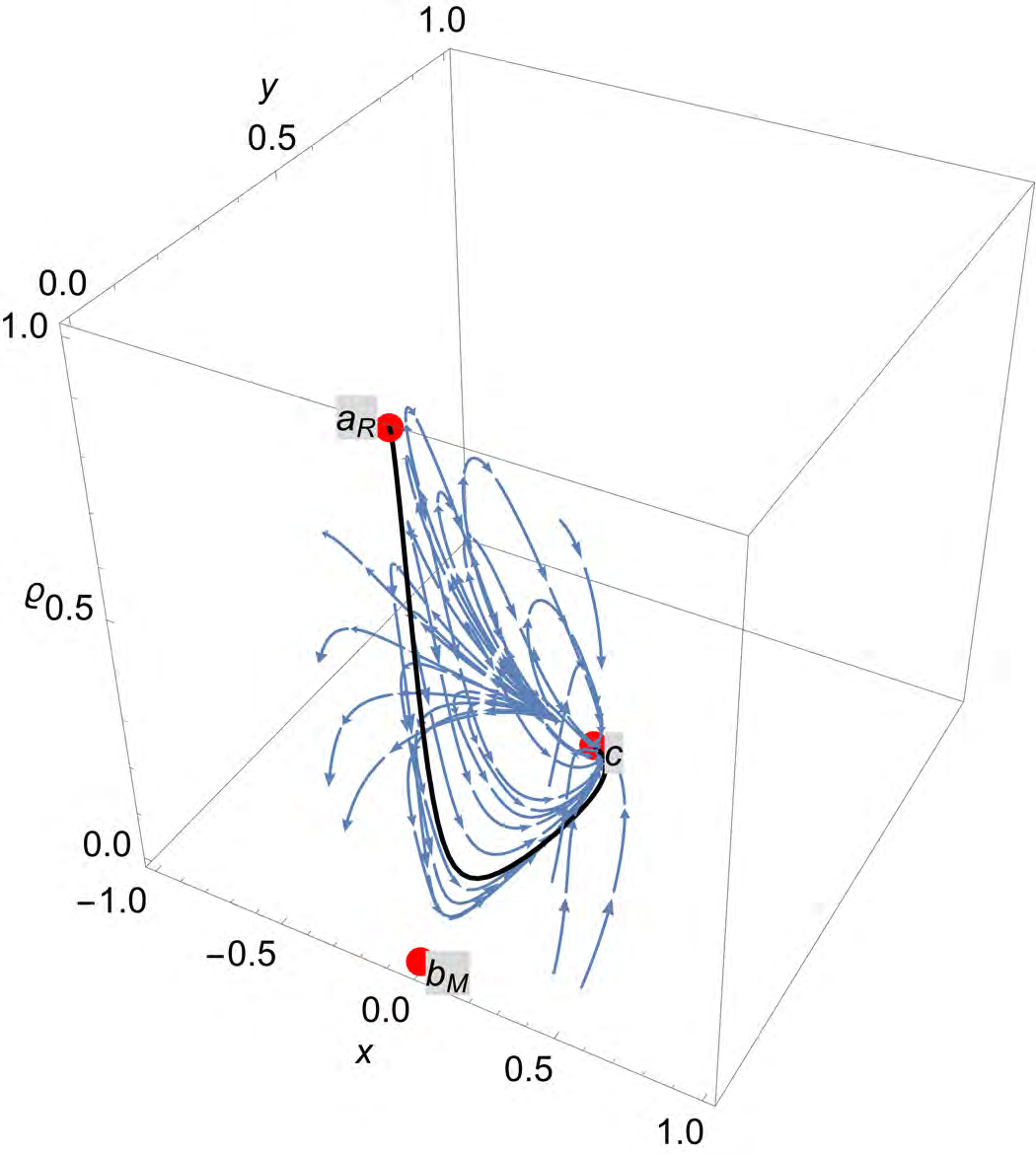}
        \caption{\scriptsize The phase space for \(\alpha = 0.347\) is shown, where the critical points are marked as red dots with their respective labels. Each trajectory represents the evolution of the Universe for different initial conditions. In particular, the black line corresponds to a Universe initialized with \( x_i = 8.0 \times 10^{-9} \), \( y_i = 1.645 \times 10^{-3} \), and \( \varrho_i = 0.999825 \).}  
    \label{3Dphase1}
\end{figure}

{
On the other hand, we conducted a numerical analysis of the autonomous system described by Eqs. ~\eqref{dinsyseq1} - \eqref{dinsyseq3}. We aim to evaluate how effectively our model captures the three states of our Universe: a radiation-dominated era, a saddle point representing a combination of dark sector components, and a final attractor for the accelerated expansion of our Universe, and to compare its predictions with recent observational data on cosmological parameters. 
Figure \ref{2Dphase1} depicts the evolution of the phase space for the trajectory $ a_R \to b_M \to c $ in 2D. The plot shows \( x \) versus \( y \) when \(\alpha = 0.347\), with critical points indicated by red dots and labeled accordingly. The different trajectories reflect various possible evolutionary paths of the Universe, depending on the initial conditions. Each blue line corresponds to a distinct set of initial conditions. In particular, the black trajectory starts from the initial values \( x_i = 8.0 \times 10^{-9} \), \( y_i = 1.645 \times 10^{-3} \), and \( \varrho_i = 0.999825 \). The fact that all trajectories converge to the same late–time solution clearly demonstrates the attractor nature of the critical point $c$, in agreement with the stability conditions derived analytically. Regarding the green-shaded region, it represents only the domain of accelerated expansion within the phase space. Its relatively small size reflects the fact that acceleration is realized only in a restricted portion of the parameter space, but importantly, our late–time attractor lies precisely inside this region. This confirms that the model consistently evolves toward accelerated expansion while satisfying the physical constraints.
Similarly, Figure \ref{3Dphase1} presents the phase-space stream flow of the trajectory \( a_R \to b_M \to c \) in 3D. This representation considers the entire phase space structure for \(\alpha = 0.347\), with critical points again marked as red dots and labeled accordingly. The black trajectory follows the same initial conditions as specified in the two-dimensional case. Both figures demonstrate that the solutions of the autonomous system asymptotically approach the attractor \( c \). Numerical simulations confirm that the proposed model effectively accounts for a late-time acceleration phase consistent with observational constraints.
}

\begin{table}[H]
 \centering
 \caption{Critical points for the autonomous system. }
\begin{center}
\begin{tabular}{c c c c c c c c c}\hline\hline
Name &  $x_c$ & $y_c$ &  $\varrho_{c}$  \\\hline
$\ \ \ \ \ \ \ \ a_{R} \ \ \ \ \ \ \ \ $ & $0$ & $0$& $1$ \\
$\ \ \ \ \ \ \ \ b_{M} \ \ \ \ \ \ \ \ $ & $0$ & $0$ & $0$ \\
$\ \ \ \ \ \ \ \ c \ \ \ \ \ \ \ \ $ & $0$ & $\frac{1}{\sqrt{3}}$ & $0$ \\

\\ \hline\hline
\end{tabular}
\end{center}
\label{table1}
\end{table}

\begin{table}[H]
 \centering
 \caption{Cosmological parameters for the critical points in Table \ref{table1}. }
\begin{center}
\begin{tabular}{c c c c c c}\hline\hline
Name &   $\Omega_{de}$ & $\Omega_{m}$ & $\Omega_{r}$ & $\omega_{de}$ & $\omega_{tot}$ \\\hline
$a_{R}$ & $0$ & $0$ & $1$ & $\frac{1}{3} \left(8 \alpha ^2-4 \alpha +1\right)$ & $\frac{1}{3}$ \\
$b_{M}$ & $0$ & $1$ & $0$ & $\frac{1}{3} \left(12 \alpha ^2-8 \alpha +1\right)$ & $0$ \\
$c$ & $1$ & $0$ & $0$ & $\frac{1-6 \alpha }{3-6 \alpha }$ & $\frac{1-6 \alpha }{3-6 \alpha }$
\\ \hline\hline
\end{tabular}
\end{center}
\label{table2}
\end{table}

In the next section, we will investigate the stability of these critical points using a linear perturbation analysis of the dynamical variables.


\textbf{Stability of critical points}\\

Now, we introduce small perturbations around each critical point and analyze how these perturbations enable the linearization of the system's equations, \( x = x_c + \delta x \), \( y = y_c + \delta y \), and \( \varrho = \varrho_c + \delta \varrho \). This procedure leads to the construction of the perturbation matrix \(\mathcal{M}\), where the eigenvalues \(\mu_1\), \(\mu_2\), and \(\mu_3\) play a crucial role in determining stability. The classification of stability generally includes the following cases:
\begin{itemize}
    \item[(i)] A stable node, characterized by all negative eigenvalues.
    \item[(ii)] An unstable node, identified by all positive eigenvalues.
    \item[(iii)] A saddle point, which exhibits a mix of positive and negative eigenvalues.
    \item[(iv)] A stable spiral, which has a negative determinant and eigenvalues with negative real parts.
\end{itemize}

Attractor points, such as stable nodes or spirals, arise in cosmic evolution regardless of initial conditions, provided that they lie within the attraction basin. Following this approach, the eigenvalues and stability conditions for each critical point are presented in detail below.

\begin{itemize}
    \item Point $a_R$ has the eigenvalues
    \begin{equation}
        \mu_1 = 1, \ \ \ \ \ \mu_2 = \frac{1}{2} \left(-\sqrt{8 \alpha +1}-1\right), \ \ \ \ \ \mu_3 = \frac{1}{2} \left(\sqrt{8 \alpha +1}-1\right), 
    \end{equation}
    therefore, it is a saddle point.
    \item Point $b_M$ has the eigenvalues
    \begin{equation}
        \mu_1 = -\dfrac{1}{2}, \ \ \ \ \ \mu_2 = \frac{1}{4} \left(-\sqrt{48 \alpha +1}-3\right), \ \ \ \ \ \mu_3 = \frac{1}{4} \left(\sqrt{48 \alpha +1}-3\right), 
    \end{equation}
    where, it is a saddle point for $\alpha >\frac{1}{6}$.
    \item Point $c$ has the eigenvalues
    \begin{equation}
        \mu_1 = \frac{1}{2 \alpha -1}, \ \ \ \ \ \mu_2 = \frac{2 \alpha }{2 \alpha -1}, \ \ \ \ \ \mu_3 = \frac{6 \alpha -1}{2 \alpha -1}, 
    \end{equation}
    which, it is stable for $1/6<\alpha <1/2$. {This bound will be of great help in determining the exact value of the parameter $\alpha$ by implementing the $\chi^{2}$ method in conjunction with observational data on supernovae. }
\end{itemize}
{It is also worth clarifying the special cases $\alpha=0$ and $\alpha=1/6$. 
For $\alpha=0$, the vector sector effectively reduces to the 
Maxwell theory, mimicking dark radiation, while for $\alpha=1/6$, the system 
behaves as a matter-like component dominated by kinetic energy. However, both cases lie at the boundary of the stability region and are excluded once 
the condition $1/6 < \alpha < 1/2$ is imposed from the eigenvalue analysis. 
In particular, the value $\alpha=1/6$ corresponds to a marginal (metastable) 
situation, while the $\chi^2$ fit with supernovae data selects a best-fit value 
of $\alpha=0.347$, well inside the stable interval. Therefore, the physically 
relevant regime of the model does not suffer from these pathologies.
}

\subsection{SN data}

As usual, we rely on $\chi^2$ minimization to obtain the numerical value of the parameter $\alpha$ that best fits the data. In this work, we use Supernovae Ia, which serve as standard candles and provide measurements of the luminosity distance \(D_L(z)\) through the observed distance modulus
\begin{equation}
\mu_{\rm obs} = m_B^{\rm corr} - M_B,
\end{equation}
where \(m_B^{\rm corr}\) is the corrected apparent magnitude and \(M_B\) denotes the absolute magnitude of the source.  
The theoretical prediction for this cosmological model with parameters \(\mathbf{p} = \{ H_0, \Omega_m, \alpha \}\) is
\begin{equation}
\mu_{\rm th}(z_i) = 5\log_{10}\!\left[\frac{D_L(z_i;\mathbf{p})}{10\,{\rm pc}}\right] + 25,
\qquad
D_L(z)=c(1+z)\int_0^z \frac{dz'}{H(z';\mathbf{p})},
\end{equation}
where the luminosity distance is expressed in terms of the background expansion rate \(H(z)\).  

In particular, we use the most recent Pantheon+ compilation together with the SH0ES Cepheid-calibrated subsample, which jointly constrain both the Hubble-flow SNe Ia and the subset of nearby SNe whose absolute magnitudes are anchored to the geometric distances inferred from Cepheid variables.  
The inclusion of SH0ES plays a crucial role in breaking the degeneracy between \(M_B\) and \(H_0\), since the calibrator SNe provide an external distance scale that propagates through the distance ladder.  
As described in Refs.~\cite{Brownsberger:2021uue,Popovic:2021yuo,Carr:2021lcj,Peterson:2021hel,Brout:2021mpj,Scolnic:2021amr,Brout:2022vxf,Yuan_2022,Riess:2021jrx}, the Pantheon+SH0ES dataset incorporates a detailed modeling of all statistical uncertainties, light-curve standardization corrections, zero-point calibration effects, host-galaxy correlations, and the full SH0ES covariance arising from Cepheid period–luminosity relations and anchor distances.  

The chi-square associated with the full Pantheon+SH0ES supernova sample is computed as
\begin{equation}
\chi^2_{\rm SN} =
(\boldsymbol{\mu}_{\rm obs}-\boldsymbol{\mu}_{\rm th})^{T}
\, \mathbf{C}^{-1} \,
(\boldsymbol{\mu}_{\rm obs}-\boldsymbol{\mu}_{\rm th}),
\end{equation}
where \(\boldsymbol{\mu}_{\rm obs}\) contains the distance moduli of both the Hubble-flow SNe and the calibrated SH0ES SNe, and \(\mathbf{C}\) is the full covariance matrix provided by the collaboration.\footnote{Pantheon+SH0ES data and covariance matrices are publicly available at \url{https://github.com/PantheonPlusSH0ES}.} Where, it is important to note that \(\mathbf{C}\) contains not only the usual statistical and systematic contributions associated with the Hubble-flow SNe, but also the correlations induced by the Cepheid calibration and the internal SH0ES error budget. Therefore, enabling robust constraints on \(M_B\), and the cosmological parameters \(\mathbf{p}\). For this model, we found the next mean values and their corresponding $1\sigma$ uncertainties: $H_0 = 66.23^{+5.67}_{-5.50}$, 
$\Omega_m = 0.33815^{+0.03406}_{-0.04085}$, 
$M_B = -19.24282^{+0.03107}_{-0.03032}$, 
$\alpha = 0.34706^{+0.00065}_{-0.00060}$.

\begin{figure}[!h]
    \centering
        \includegraphics[scale=0.5]{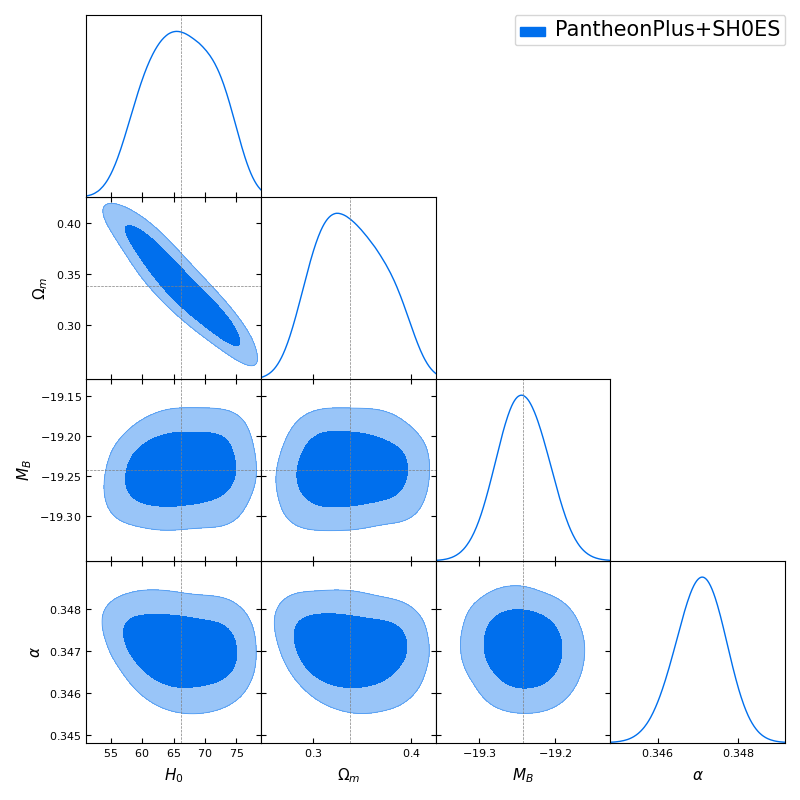}
        \caption{\scriptsize Confidence levels at the $1\sigma$ and $2\sigma$ limits for the Bumblebee model obtained from the PantheonPlus+SH0ES dataset. The contours show the marginalized constraints on $(H_0,\ \Omega_m,\ M_B,\ \alpha)$. }  
    \label{contours_chi2}
\end{figure}

\begin{figure}[H]
    \centering
        \includegraphics[scale=0.57]{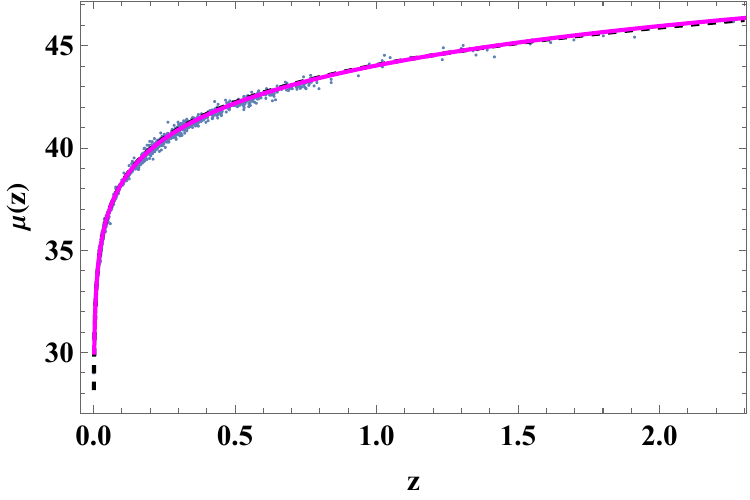} \
        \caption{Distance modulus versus redshift. Shown are: Data points, Bumblebee model (solid curve), and $\Lambda$CDM model (dashed curve). 
        }  
    \label{fig:3}
\end{figure}

\begin{figure}[H]
    \centering
        \includegraphics[scale=0.57]{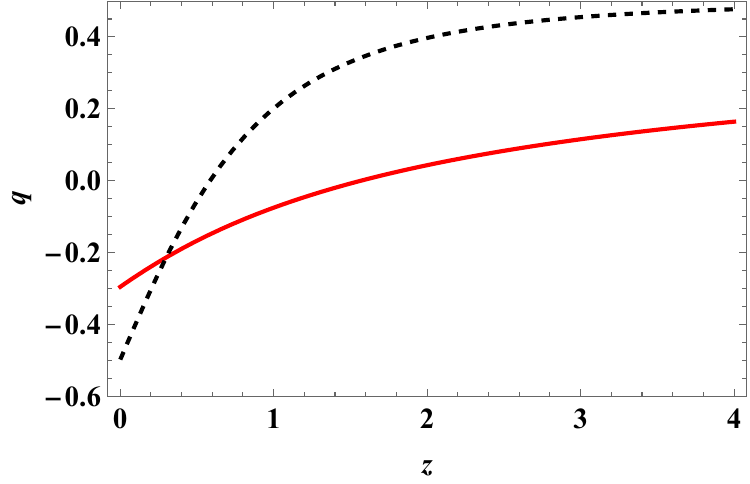} \
        \includegraphics[scale=0.57]{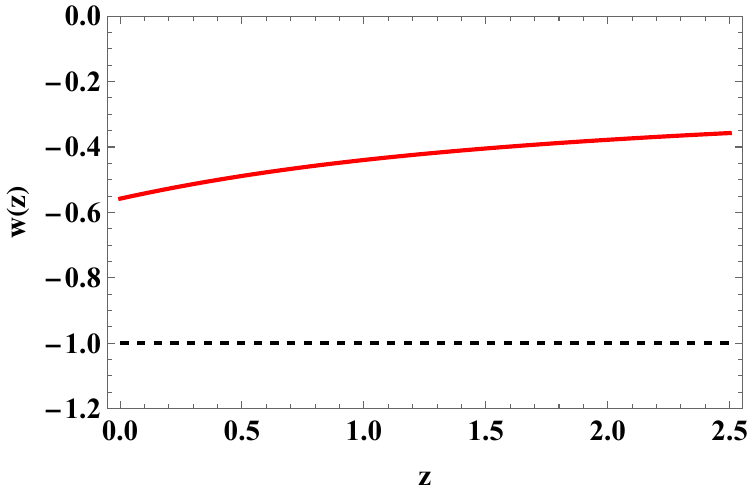}
        \caption{Deceleration parameter (left panel) and dark energy equation-of-state parameter (right panel) versus redshift. In both panels, the dashed curves correspond to the $\Lambda$CDM model.}  
    \label{fig:4}
\end{figure}

\begin{figure}[H]
    \centering
        \includegraphics[scale=0.8]{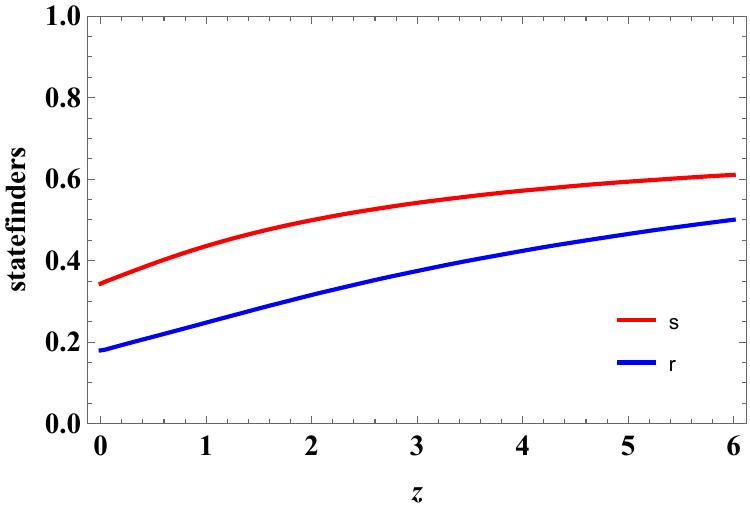} 
        \caption{Statefinder diagnostic parameters $r$ (in blue) and $s$ (in red) versus redshift.}  
    \label{fig:5}
\end{figure}

\begin{figure}[H]
    \centering
        \includegraphics[scale=0.7]{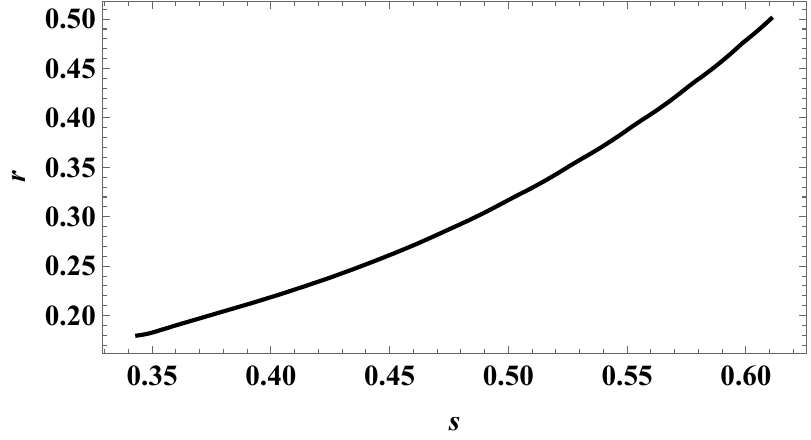} 
        \caption{Parametric plot $r$ versus $s$.}  
    \label{fig:6}
\end{figure}

Next, for this value of the parameter $\alpha$, we compute the distance modulus, the fractional densities, the decelerating parameter, the DE equation-of-state parameter as well as the statefinders versus redshift, which are shown in Figures \ref{fig:3}, \ref{fig:4}, \ref{fig:5} and \ref{fig:6}.


For the sake of comparison, we show in the same plot the curves corresponding to the $\Lambda$CDM model as well (black dashed curves). We see that the distance modulus fits the SN data nicely, and it is indistinguishable from the concordance model. Our results indicate that the Universe passes from a decelerating phase to an accelerating one at late times (or at low redshift, i.e., recently), albeit a bit sooner compared to the concordance cosmological model. Initially, namely at high redshift, the Universe was dominated by radiation, whereas today it is dominated by dark energy, although according to our findings $\Omega_{m,0}$ is quite small compared to the more standard value around 0.3. Furthermore, the equation-of-state parameter of dark energy always remains higher than the $-1$ line, and hence it corresponds to quintessence rather than phantom. 
{
Finally, plots \ref{fig:5} and \ref{fig:6} illustrate the evolution of the statefinder parameters \( r \) and \( s \) as functions of redshift \( z \), along with their parametric relation for the Bumblebee model with a cosmic triad. As the trajectory in the \( r \) vs. \( s \) plot evolves, it ultimately converges to the point \( (1,0) \), indicating that the model asymptotically approaches \( \Lambda \)CDM in its final stage. This suggests that a Bumblebee field using a cosmic triad is relevant but gradually diminishes, allowing the expansion dynamics to transition toward a \( \Lambda \)CDM-like behavior. The \( r \) vs. \( z \) plot shows an initial decrease, then a minimum before increasing again, which reflects a transient phase where deviations from standard cosmology are more pronounced before stabilizing. On the other hand, the \( s \) vs. \( z \) plot exhibits a decrease in \( s \), meaning that at late times, the model evolved towards  \( \Lambda \)CDM. This behavior suggests that while the Bumblebee field and the cosmic triad contribute to the dynamics at high redshift, their influence weakens as the Universe expands. The parametric \( r \)-\( s \) curve confirms that while the model follows a distinct trajectory at earlier stages, its final evolution aligns with \( \Lambda \)CDM. This result is consistent with scenarios where additional fields or modifications to the energy content of the Universe play a role in the early evolution but become subdominant at late times. The findings suggest that while present-day observations might capture effects from the Bumblebee field and the cosmic triad, the long-term fate of the Universe in this model is indistinguishable from the standard \( \Lambda \)CDM scenario.}


{Before concluding, it is useful to compare our results with previous studies. In \cite{Armendariz-Picon:2004say}, it was found that the dark-energy equation-of-state parameter may cross the phantom divide, and that the cosmological dynamics exhibits a single de Sitter attractor. In \cite{BeltranJimenez:2009xus}, observational constraints suggested that a spatially flat universe is disfavored within that specific framework.}
{
In contrast, in the scenario analyzed here, the combination of a cosmic triad configuration with a non-minimal Ricci–Bumblebee interaction leads to a richer phase-space structure. In particular, the model naturally reproduces radiation- and matter-dominated saddle points together with a late-time accelerated attractor compatible with current observational data.}


\section{Conclusions}\label{Concluding_Remarks}

In this work, we have investigated the cosmological implications of a Bumblebee gravity model, a covariant theory with spontaneous breaking of both Lorentz symmetry and diffeomorphisms and non-minimal coupling with gravitational interaction. Within this framework, elementary-level studies of the cosmological set require certain modifications for proper implementation. That is, if the goal is to work on an FLRW-type geometry (such as that observed for the universe today), it is necessary to introduce a cosmic triad and thus recover the isotropy of spacetime on large scales. In particular, this formalism has been implemented in this work to be consistent with cosmological analyses. In this context, a model of the Bumblebee family was considered, which has only one coupling term with gravity (\ref{action00}). Through dynamical systems analysis performed on the modified Friedmann equations (\ref{MFreq1})-(\ref{ii}), it was found that the parameter is bounded from below and above according to $1/6<\alpha<1/2$. This interval corresponds to the stable one for the critical points. With this information at hand and employing the $\chi^{2}$ algorithm along with SN data, the best value determined for the parameter $\alpha$ was 0.347. Taking into account this result, the phase space  in Figs. \ref{2Dphase1} and \ref{3Dphase1} show that this model is capable of describing an initial Universe dominated by matter evolving to an accelerated epoch. Furthermore, Fig. \ref{fig:3} depicts that this model is in agreement with SN data. On the other hand, Fig. \ref{fig:4} shows that this particular Bumblebee theory corresponds to an accelerated Universe dominated by a quintessence field. This fact is corroborated by Fig. \ref{fig:5}, where the statefinder parameters drift apart from those corresponding to the $\Lambda$CDM model. These results suggest that Bumblebee models can be considered as potential candidates to challenge the $\Lambda$CDM cosmological model.

{From a physical point of view, our results show that the Ricci–Bumblebee interaction can play an important role in the late-time cosmological dynamics. Even in the simplified scenario considered here, the model exhibits a rich phase-space structure including radiation- and matter-dominated saddle points together with a stable accelerated attractor. This demonstrates that spontaneous Lorentz symmetry breaking encoded in the Bumblebee field can effectively contribute to the cosmological expansion without the need for introducing an explicit cosmological constant.}
{
Moreover, the comparison with observational supernova data provides a quantitative constraint on the coupling parameter of the theory. The obtained best-fit value lies within the theoretically allowed interval derived from the stability analysis, indicating that the model is consistent with current cosmological observations. At the same time, the behavior of the statefinder diagnostics shows that the scenario can, in principle, be distinguished from the standard $\Lambda$CDM model.}

{Although the model contains only one free parameter, its origin is fundamentally different from scalar--potential scenarios such as \cite{Armendariz-Picon:2004say,BeltranJimenez:2008iye,BeltranJimenez:2009xus}. In our case, late-time acceleration arises from the anisotropies induced via the Ricci tensor non-minimally coupled to Bumblebee, rather than from an \textit{ad hoc} potential and fixed numerical values for the coupling constants. This framework, combined with the cosmic triad construction, ensures compatibility with isotropy and yields stable accelerating attractors in the dynamical system analysis. 
} 
An interesting aspect of the present scenario is that the late-time acceleration arises effectively from the Ricci–Bumblebee interaction without introducing an explicit cosmological constant. In this sense, the Bumblebee field can be interpreted as providing a geometrical source of dark energy associated with spontaneous Lorentz symmetry breaking.
This can function as an initial point for studies with more complex couplings and thus restricts the possible coupling values of the complete theory. Consequently, the viability of Bumblebee models as potential cosmological candidates to elucidate the accelerated expansion of the Universe can be substantiated or refuted. Another important open issue is the perturbative consistency of Bumblebee cosmology beyond the background level. In particular, recent no-go results obtained for the most general marginal Bumblebee action \cite{vandeBruck:2025aaa} indicate that achieving a homogeneous and isotropic FLRW background together with healthy cosmological perturbations is highly non-trivial. In this respect, the present background analysis should be viewed as a first step toward identifying phenomenologically viable reduced sectors of the theory.

\section{Acknowledgments}
M. Gonzalez-Espinoza acknowledges the financial support of FONDECYT de Postdoctorado, N° 3230801.

\section*{Conflict of Interest}

The authors declare no conflict of interest.

\section*{Data Availability Statement}

Data sharing is not applicable to this article as no new data were created or analyzed in this study.

\section*{Keywords}

Cosmology; Bumblebee theory; Dark energy; Observational data.

\bibliography{bio}

\end{document}